\documentclass[aps,prl,reprint,nofootinbib,floatfix]{revtex4-2}
\usepackage[T1]{fontenc}
\usepackage{amsmath,amssymb,bm,graphicx}
\usepackage[colorlinks=true,citecolor=blue,urlcolor=black,linkcolor=blue]{hyperref}
\hypersetup{pdftitle={Classical records and Bell correlations in a dynamically formed quantum body},pdfauthor={XinXin Qi and Hao Sun}}
\newcommand{\Tr}{\operatorname{Tr}}
\newcommand{\col}{\mathrm{col}}
\newcommand{\dd}{\,\mathrm d}
\begin{document}
\title{Classical records and Bell correlations in a dynamically formed quantum body}
\author{XinXin Qi}
\email{qxx@dlut.edu.cn}
\affiliation{Institute of Theoretical Physics, School of Physics, Dalian
University of Technology, No. 2 Linggong Road, Dalian, Liaoning 116024,
P.R. China}
\author{Hao Sun}
\email{haosun@dlut.edu.cn}
\affiliation{Institute of Theoretical Physics, School of Physics, Dalian
University of Technology, No. 2 Linggong Road, Dalian, Liaoning 116024,
P.R. China}

\begin{abstract}
Competing quantum contacts can create a boundary that was absent from the specification of a subsystem. We construct approximately isolated bodies from bare field vacuum and zero-centred contact packets. A binding contribution with vanishing mean energy density nevertheless changes a bounded formation signal by order one at the same clock. On the formed material spins and their original memories, coarse collective records precede an all-outcome Bell violation. The field evolves dynamically, with the finite bath, residual boundary interaction, failed records and preparation costs retained. The result connects dynamical subsystem formation, classical collective records and microscopic nonlocality, with polynomial resources in the stated preparation and measurement class.
\end{abstract}
\maketitle

When two objects bind, they become less sensitive to some of the degrees of freedom around them. The boundary used in a subsequent calculation is then a consequence of the earlier motion. In a quantum system, this leaves a further question: does the formation of an approximately isolated body also remove the correlations that a sufficiently fine measurement could reveal?

We study contacts that compete for the same endpoints. A growing contact lowers the energy of a coupled bosonic field and suppresses neighboring contacts. The resulting boundary supports a short experiment on the selected material spins. First, collective detectors read their means with a resolution broader than the fluctuations of an average over many labels. Then local detectors resolve a Bell correlation on those same labels. The coarse records and Bell bits belong to one physical sequence.

Coarse measurements can give classical collective statistics~\cite{r1,r9,kofler}, while weak measurements can preserve nonlocality for later observers~\cite{silva}. Those studies take the measured system as given. Here contact competition supplies its boundary, whose error is controlled through the joint record and Bell program. Dynamical adjacency~\cite{r11} and spin transport~\cite{r10,r15} provide related mechanisms. The theory and principal comparisons appear in the companion article~\cite{paperB}; the assumptions and estimates used here are collected in Appendices~A--F.

\textit{A boundary produced by contact competition.---}
There are two classes of $n$ sites and $N=n^2$ possible contacts. Each contact has a coordinate $q_{ij}$ and a weight $F(q_{ij})=\sin^2(\pi q_{ij}/2)$. With $x_{ij}=q_{ij}^2$, $r_i=\sum_jx_{ij}$ and $c_j=\sum_ix_{ij}$, the contact potential is
\begin{equation}\label{eq:U}
U=\sum_i(r_i-1)^2+\sum_j(c_j-1)^2
 +\sum_{ij}x_{ij}(x_{ij}-1)^2.
\end{equation}
Its negative quadratic term starts the growth. Positive sextic terms limit a contact's amplitude, while positive quartic terms penalize contacts sharing a site. The two site classes and the valence preference are given; no edge is designated as the winner.

\begin{figure*}[t]
\centering
\includegraphics[width=\textwidth]{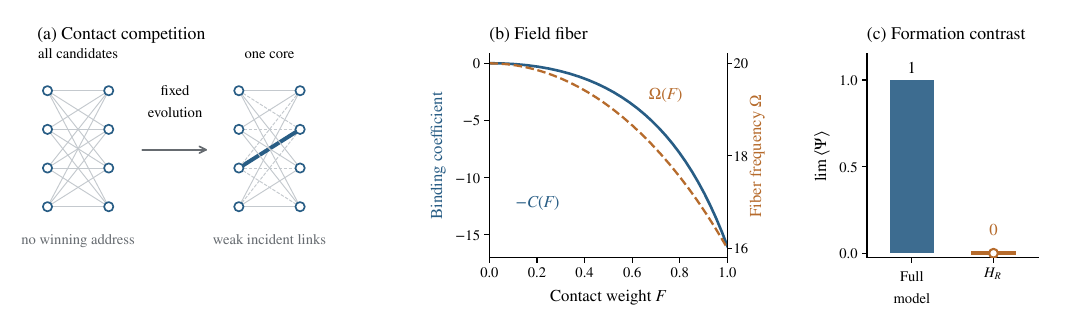}
\caption{\label{fig:formation}Contact competition and binding. (a) All bipartite contacts are initially available; the drawing illustrates one core with weak incident links. (b) Exact fixed-coupling binding coefficient and field-fibre frequency; the energy shift is $-MC-Z$. These curves specify the Hamiltonian, not a prepared ground state or trajectory. (c) The same-clock hot-amplitude test in Eq.~\eqref{eq:surface} tends to one in the physical model and zero after deleting the complete binding surface. The bars are limits, not finite-size samples or core counts. }
\end{figure*}

Two bosonic modes on each contact supply an additional binding surface. In the nominal filled-source sector their exact fixed-$q$ decomposition is
\begin{equation}\label{eq:binding}
h_{\mathrm{bare},e}=h_{\mathrm{exc},e}-MC(F_e)-Z(F_e),\qquad
h_{\mathrm{exc},e}\ge0,
\end{equation}
where $C(f)=128f^2/(20-12f)$ and $Z(f)=20-\sqrt{400-144f^2}$. The fibre excitation frequency lies between 16 and 20. The binding becomes deeper as the contact strengthens [Fig.~\ref{fig:formation}(b)]. The physical input is a bare field vacuum with zero-centred contact packets. Controlling its actual propagation is therefore necessary before Eq.~\eqref{eq:binding} can be used to infer formation~\cite{r2}.

The fixed scaling is $M=n^{16}$, contact mass $m=1+M^3$, and bath length $L=8Mn^3$. At the preset formation clock
\begin{equation}\label{eq:clock}
\tau_f=\frac{Mn}{(3/4)g_n(3\log n+C_T)},\qquad
g_n=16+\frac{16\pi^4}{5}+\frac{9\pi^4}{5M},
\end{equation}
the unmonitored commuting-$q$ Born law contains at least
$\lfloor c\sqrt{n/\log n}\rfloor$ disjoint cores with probability tending to one. A core has one contact near $|q|=1$ and the other contacts incident to its two endpoints near zero, to accuracy $a=n^{-11/2}$. The proof combines actual dressed-field propagation, a bounded endpoint comparison, and nonlinear entrance, competition and cooling estimates~\cite{paperB}; see Appendices~A--F. It does not select a classical path or require all sites to form a matching.

Binding changes formation even when its mean energy density vanishes. At the same bare input and clock, compare the filled-sector world with $H_R=H+\Delta H$, where $\Delta H=M\sum_eC(F_e)+\sum_eZ(F_e)$. For the bounded hot-amplitude test $\Psi=\operatorname{clip}_{[0,1]}(4\max_e|q_e|-1)$,
\begin{equation}\label{eq:surface}
\begin{aligned}
\langle\Psi\rangle_H-\langle\Psi\rangle_R&\longrightarrow1,\\
\frac{\langle\Delta H\rangle_H+\langle\Delta H\rangle_R}{Mn}&\longrightarrow0.
\end{aligned}
\end{equation}
The quartic onset of $C$ and $Z$ reinforces larger seeds in the weak cloud. Deleting this subextensive surface suppresses strong-contact formation at the original clock. Both worlds retain the fields and finite bath; each energy mean uses its own state. The comparison isolates the surface retained in correct Born--Oppenheimer dynamics~\cite{r2,paperB}; $\Psi$ tests amplitude, while the preceding theorem supplies the core count.

A finite native readout locates a guarded core. A failed address receives a declared fallback or default program and remains in the sample. In the frozen reference, quadratic flatness of $F$ gives total external weight
\begin{equation}\label{eq:boundary}
b_{\mathrm{ext}}=\sum_{\mathrm{external,\ incident}}F(z_e)
 \le Cna^2=Cn^{-10}.
\end{equation}
The sum matters: small individual links alone would not isolate the body. For $K=n^2$ spin labels the generator error is $O(n^{-8})$; over the allowed late duration at most $n^4$ it contributes $O(n^{-4})$. The comparison from the evolving quantum boundary to this frozen reference has its own additional error.

\textit{A collective record that leaves the spins available.---}
Every site carries $K$ material spins $\bm\sigma$ and $K$ original-memory spins $\bm\tau$. The memory exchange has strength one from physical time zero. In a selected body, one label has generator
\begin{equation}\label{eq:hB}
h_B=J\bm\sigma_i\!\cdot\!\bm\sigma_j+
\bm\sigma_i\!\cdot\!\bm\tau_i+
\bm\sigma_j\!\cdot\!\bm\tau_j,\qquad J>0.
\end{equation}

For each label independently, a uniform classical phase chooses one of
$0,\pi/2,\pi,3\pi/2$; all $4n$ spins of that label begin in its corresponding equatorial product state. No Bell pair is supplied. This input is fully separable and is separated from the fixed Jordan--Wigner even-quasifree class. Its exchange symmetry makes the forming spin feedback exactly $JK\sum_eF(q_e)+2nK$. The original memories are exchange-dark in this particular preparation until a right-material $Y$ pulse activates their response.

After that pulse, a fixed finite set of collective queries measures
$\overline A_j=K^{-1}\sum_\ell A_{j,\ell}$, from the declared four-spin Pauli-tensor library. Query times are preset and separated, before $t_*/2$. Their scalar reference centres follow
\begin{equation}\label{eq:rho}
\rho_4(t)=e^{-ith_B}U_Y\rho_4U_Y^\dagger e^{ith_B},\quad
\rho_4=\frac14\sum_\phi(|+\phi\rangle\langle+\phi|)^{\otimes4},
\end{equation}
with $U_Y=e^{-i\theta\sigma_j^Y/2}$ and $\theta=\pi/6$.

The collective detector has decoded width
$K^{-1/2}\ll\sigma=n^{-1/16}\ll1$. It resolves the mean increasingly well while remaining broad compared with the fluctuations of one average. In each scalar-prefix reference, $\operatorname{Var}\overline A_j\le1/K$. Comparing the full positive query squares
$(P_j+g_j\overline A_j)^2/2$ and
$(P_j+g_j\mu_j)^2/2$ gives a per-query label term
$O((\sigma\sqrt K)^{-1})=O(n^{-15/16})$.
Clock, world, readout and preparation terms are additional.

This comparison keeps the material spins, memories, physical pointers and prescribed quantum references. Subsequent Bell instruments therefore act on the output of the same comparison. The classical description is the density flow of Eq.~\eqref{eq:rho} with noisy collective readout; it describes a specified finite query history. The density can contain internal entanglement. Neither the detector resolution nor the number of copies is created by nucleation. The coarse resolution does not relax calibration requirements for the subsequent Bell analyzers~\cite{jeong}.

\begin{figure*}[t]
\centering
\includegraphics[width=\textwidth]{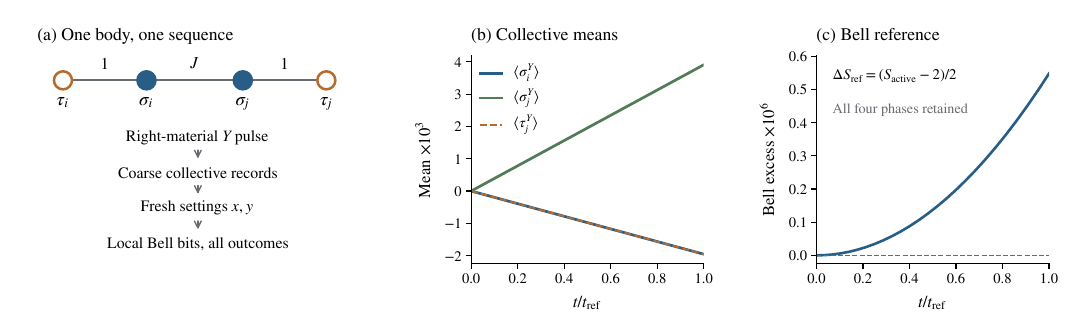}
\caption{\label{fig:records}Two resolutions on the same material. (a) A right-material pulse precedes coarse records, then fresh Bell settings; all $K$ labels remain. (b) Exact isolated four-spin means for $J=1$, $\theta=\pi/6$. (c) All-phase Bell excess; analyzers are recalibrated at each plotted time. The reference clock is $t_{\mathrm{ref}}=J/[64(1+J)^2]$, while the physical protocol uses one $t_*=\min(T_0/2,t_{\mathrm{ref}})$ and its fixed analyzers in the allowed $T_0>0$ window.}
\end{figure*}

\textit{Fine correlations after the coarse measurements.---}
At body time $t_*=\min\{T_0/2,J/[64(1+J)^2]\}$, Alice and Bob receive one global pair of independent uniform settings $x,y$, independent of all past records and quantum systems. In the two active preparation phases, the exact one-flip motion on the path $(\tau_i,\sigma_i,\sigma_j,\tau_j)$ with weights $(1,J,1)$~\cite{r15} gives
\begin{equation}\label{eq:S}
\begin{aligned}S_{\mathrm{active}}(t_*)&=
2\sqrt{1+4\sin^4(\theta/2)P_A(t_*)P_B(t_*)}\\&=2+\eta_B>2.\end{aligned}
\end{equation}
Here $P_A,P_B$ are normalized left and right masses within that one-flip sector. The other two phases give the declared $+1$ defaults and remain in the average.

Each party returns $K$ bits. The retained score is
\begin{equation}\label{eq:V}
V=4(-1)^{xy}K^{-1}\sum_{\ell=1}^K a_\ell b_\ell,\qquad |V|\le4.
\end{equation}
Its local bound is two, including arbitrary processing across labels inside either local bank. If $p_s$ is the probability of the actual guarded branch and $\varepsilon_{\mathrm{joint}}$ is the complete output error, then
\begin{equation}\label{eq:alltrial}
\begin{aligned}\mathbb EV&\ge(2+\eta_B/2)p_s-4(1-p_s)-4\varepsilon_{\mathrm{joint}},\\\liminf_n\mathbb EV&\ge2+\eta_B/2.\end{aligned}
\end{equation}
No phase, address, vacuum or apparatus flag removes a trial. The local-channel estimate pays for residual interaction while the instruments run. It is not a spacelike or device-independent implementation.

The finite Rubin bath~\cite{r5} is part of the forming world throughout. The observation interval used here lies well before its first return at $2(L+1)$; initial slip, free noise and nonzero image remainders are retained in the short-time comparison. A separate experiment on fresh preparations resolves the later return through a bounded momentum score~\cite{paperC}. That experiment tests the temporal limits of the same finite environment and is not appended to this Bell history.

The same quantum body thus supports smooth collective records and a microscopic Bell witness within a finite observation window. The construction uses specified inputs, measurement supplies and ideal constitutive couplings, with large polynomial costs. It establishes a dynamical existence result; autonomous machine fabrication and a classical description of every microscopic observable remain outside its scope.

\textit{Data and AI use.---}
OpenAI Codex (GPT-6 family) assisted with   analytical derivations, model construction, symbolic and numerical verification code, and manuscript editing. The authors checked the code
 and numerical results and take responsibility for
the scientific conclusions.

\clearpage
\onecolumngrid
\appendix
\setcounter{equation}{0}
\renewcommand{\theequation}{A\arabic{equation}}
\section*{Appendix A: The specified world and bare input}
There are $N=n^2$ contact coordinates and $K=n^2$ labels of material and original-memory spins at each site. With $r=n^8$, $M=r^2$, $m=1+M^3$, $L=8Mn^3$ and fixed $J>0$, the physical Hamiltonian is
\begin{align}
H={}&T_q+MU+H_B+\sum_eh_{\mathrm{bare},e}+H_{\mathrm{mat}}+V_{\mathrm{mem}}
+3JNK+6nK+17N\widehat D^2+20N,\\
T_q={}&\sum_ep_e^2/(2m),\qquad
V_{\mathrm{mem}}=\sum_{v,\ell}\bm\sigma_{v,\ell}\!\cdot\!\bm\tau_{v,\ell},\\
H_{\mathrm{mat}}={}&J\sum_{ij,\ell}F(q_{ij})\bm\sigma_{i,\ell}\!\cdot\!\bm\sigma_{j,\ell}.
\end{align}
Here $U$ is the valence/competition potential in the main text, $F(q)=\sin^2(\pi q/2)$ and
\begin{align}
h_{\mathrm{bare},e}={}&20(N_{a,e}+N_{b,e})-12F_e(a_eb_e+a_e^\dagger b_e^\dagger)
-8\widehat D F_e(a_e+a_e^\dagger+b_e+b_e^\dagger),\\
H_{B,e}={}&\frac12\sum_{j=0}^{L-1}P_j^2+\frac12(X_0-Mq_e)^2
+\frac12\sum_{j=0}^{L-2}(X_{j+1}-X_j)^2+\frac12X_{L-1}^2.
\end{align}
The nominal model has one colour, profile zero, far-end load one and a filled source $\widehat D=r$. Its mass, contact coefficient and bath geometry stay fixed in wrong-charge sectors. Every fixed-charge source term is scalar on that block, while coherent blocks retain relative phases. Charging is isolated; subsequent world couplings conserve $\widehat D$. All preparation/source errors and their own energy certificates remain part of the program.

The initial contacts are real Gaussian packets of variance $1/(2M^2)$, the fields are in bare vacuum, and the initial free-bath marginal has inverse temperature one. The active-world factorization and allowed bath--$E$ and passive-reference extensions are those of the theory article. An explicit finite-resource witness may take the passive reference trivial. General normal inputs require their own preparation cost and are not assigned the explicit Bell law by implication.

Each label independently chooses a uniform phase $\phi_\ell$ from four equatorial directions, stored in $C_\ell$. All $4n$ spins of that label begin in the corresponding product state. The spin marginal is separable and has distance at least $(\sqrt7-2)/12$ from the fixed Jordan--Wigner even-quasifree class. The certificate follows from a four-annihilator expectation $1/16$ with vanishing anomalous two-point averages; it does not exclude the class's convex hull or itself prove entanglement.

On this symmetric input the exact spin feedback during unmonitored formation is $JK\sum_eF(q_e)+2nK$. The original memories are exchange-dark then, but the unit coupling remains present and becomes active after the local material pulse. No winning edge is specified by the preparation.

\setcounter{equation}{0}
\renewcommand{\theequation}{B\arabic{equation}}
\section*{Appendix B: Formation from actual field propagation}
For $f=F(q)$, define
\begin{equation}
C(f)=\frac{128f^2}{20-12f},\qquad
Z(f)=20-\sqrt{400-144f^2},\qquad\Omega(f)=\sqrt{400-144f^2}\in[16,20].
\end{equation}
The filled-sector field split is a single-contact identity $h_{\mathrm{bare},e}=h_{\mathrm{exc},e}-MC(F_e)-Z(F_e)$; its total excitation is $h_{\mathrm{exc}}=\sum_eh_{\mathrm{exc},e}$. The field starts bare, so the fixed-$q$ identity is followed by a dynamical estimate. For the same reduced initial contact/bath/spin factor, the actual physical state and quantum mass reference satisfy
\begin{equation}
\sup_{0\le t\le T_{\max}}
\|\rho_p(t)-V_DV_P\sigma_m(t)V_P^\dagger V_D^\dagger\|_1
\le Cn^{-1/2}\log n,\qquad T_{\max}=Mn\log n.
\end{equation}
The reference retains a variable contact metric, Born--Huang term, finite bath, original spins and allowed quantum references. This is a quantum-to-quantum comparison. Its finite third-order dressing, rectangular row and two closed spectral buffers use the actual input's own high-power and force graphs. A scalar Gaussian symbol check or an assumed effective-state moment cannot replace them. The longer first-return $n^{-13/4}$ bridge is not used here.

On this shorter horizon, bounded contact endpoints can be compared with the mathematical process
\begin{align}
\dd Q&=\alpha_m\Pi\dd\tau,\\
\dd\Pi&=(-\alpha_m\Pi-\nabla V(Q))\dd\tau+\sqrt{2/M}\dd B,\\
\alpha_m&=M^3/m,\quad V=U-\sum C-\sum Z/M+s/M,\quad t=M\tau.
\end{align}
Its initial Gaussian law is the Husimi law of the exact packet, not a new physical ensemble. The scalar positive route traces $E$ before using its auxiliary bath; it does not give a positive conditional state of the full quantum environment. The physical-$E$ comparison remains separate. The endpoint error includes band, meter, force-tail, Moyal/periodic, metric, cut-exit, initial covariance/profile and nonzero finite-image terms, with leading $Cn^{-1/2}\log n$.

The nonlinear entrance produces past seeds $Y_e=A_e^2$, $A_e\sim\mathcal N(0,1)$, with $Q_e^2=(3/4n)Y_e[1+o(1)]$ at the entrance. The dense-cloud fourth moment and a common stop control pressure, active parents, cooling and noise. With $g_n=16+16\pi^4/5+9\pi^4/(5M)$, the first-wave clock is
\begin{equation}
\tau_f=Mn/[(3/4)g_n(3\log n+C_T)].
\end{equation}
Strict seed margins transfer through a finite $q$ meter to the preinstrument Born law, giving at least $\lfloor c\sqrt{n/\log n}\rfloor$ disjoint cores with probability tending to one. A core has $\big||q_e|-1\big|\le a$ and $|q_f|\le a$ on every other incident edge, $a=n^{-11/2}$. No momentum-success, CPU or field-test condition defines this unmonitored count. Later momentum outcomes are summed by trace preservation in the endpoint proof. All sites need not form a matching, and no common quantum path is inferred.

\setcounter{equation}{0}
\renewcommand{\theequation}{C\arabic{equation}}
\section*{Appendix C: A recorded boundary and the retained-spin reference}
The physical native sequence reads raw momentum and then raw position, saves finite words and computes a guarded address with fixed padding. The mathematical $q$-first endpoint meter used above is a proof tool, not a reordered actual protocol. All invalid/overflow words, dirty helpers and actual quantum failure feedback remain. A false core flag uses a declared fixed fallback.

Call the complete native-output state $A_s$. The same physical finite-copy channels act on a comparison input to construct $A_r$ on the same output Hilbert space. This full-spin reference has its own energy and spread,
\begin{equation}
\mathbb E_W(A_r)\le Cn^{21},\qquad
\|(q-z)Y_{r,V}\|_\col\le Cn^{-10}.
\end{equation}
Its spread uses position width $n^{-11}$, raw momentum scale $n^{24}$ and digit step $n^{-12}$ in the reference's joint-spectrum calculation. It is not transferred to an actual conditional state through trace distance.

For a correctly guarded word, flatness of $F$ gives
\begin{equation}
b_{\mathrm{ext}}\le Cna^2=Cn^{-10},\quad
|1-F(z_{ij})|\le Ca^2=Cn^{-11},\quad a=n^{-11/2}.
\end{equation}
Summing all $K=n^2$ labels gives generator difference $Cn^{-8}$ and integrated digital isolation error $Cn^{-4}$ over a late duration $\tau\le n^4$. The prior quantum freezing estimate is separate:
\begin{equation}
\varepsilon_{\mathrm{freeze}}\le C(n^{-7}\tau+n^{-21/2}\tau^2).
\end{equation}
It follows from the reference's own $q-z$ spread and speed under the common world part. New spin/App controls are the same in the two evolutions and do not access contact, field or bath variables. Their large coefficients do not multiply that world-freezing error.

Let $G$ be the actual valid-core branch and $p_s=\Tr A_{s,G}$, $p_r=\Tr A_{r,G}$. For $p_r>0$, use the normalized glue
\begin{equation}
\Omega_{\mathrm{in}}=(p_s/p_r)A_{r,G}+A_{s,G^c},\qquad
\|A_{s,G}-(p_s/p_r)A_{r,G}\|_1\le2\varepsilon_{\mathrm{init}}.
\end{equation}
For $p_r=0$, $p_s\le\varepsilon_{\mathrm{init}}$ and a specified normalized hot reference on mass $p_s$ has the same bound. No rare-state normalization or discarded run occurs. Both complete suffixes are trace preserving.

Every spin, memory, physical pointer and prescribed quantum $R$ is kept through the pulse and mean queries. The selected label generator is
\begin{equation}
h_B=J\bm\sigma_i\!\cdot\!\bm\sigma_j+\bm\sigma_i\!\cdot\!\bm\tau_i+
\bm\sigma_j\!\cdot\!\bm\tau_j.
\end{equation}
The older successful spin-tracing field-Bell construction is a different instrument and supplies no proof of this retained-spin history.

\setcounter{equation}{0}
\renewcommand{\theequation}{D\arabic{equation}}
\section*{Appendix D: The collective query and its physical copy}
The right-material pulse is $U_Y=e^{-i\theta\sigma_j^Y/2}$, $\theta=\pi/6$. For the four-phase mixture $\rho_4$, the scalar means are generated by $\rho_4(t)=e^{-ith_B}U_Y\rho_4U_Y^\dagger e^{ith_B}$. The finite Pauli-tensor query library and separated times $0<t_1<\cdots<t_m\le t_*/2$ are fixed in advance, with
\begin{equation}
t_* =\min\{T_0/2,J/[64(1+J)^2]\},\qquad d_g=n^{-40},\qquad\sigma=n^{-1/16}.
\end{equation}
$T_0>0$ is the allowed observation window. The initial density point is fixed before the future query plan or settings. It can remain internally entangled; the classical claim concerns the noisy collective history.

For $\overline A=K^{-1}\sum_\ell A_\ell$, the positive query square and its scalar counterpart differ by
\begin{equation}
gP(\overline A-\mu)+\tfrac12g^2(\overline A^2-\mu^2),\qquad \int g\dd t=d_g.
\end{equation}
In a scalar-prefix reference $\operatorname{Var}\overline A\le1/K$ and the fresh pointer has $\|PY\|=1/(2d_g\sigma)$. The square term remains. Duhamel costs
\begin{equation}
C[(\sigma\sqrt K)^{-1}+d_g/\sigma+d_g/\sqrt K+d_g^2]
\end{equation}
per query. The telescope uses actual suffixes after scalar prefixes, not product assumptions on the actual postquery state. A later full Bell channel contracts this same normal error without another factor $K$.

Each pointer is a specified chirped Gaussian focused to physical width $d_g\sigma$. Its read variable is $Q/d_g=\mu+\xi$ with Gaussian decoded width $\sigma$. Finite positive copying uses at most $Cn^2$ boundary qubits, decoded transition half-width $n^{-100}$ and duration $n^{-1000}$. The reference's own derivative graph gives
\begin{equation}
\varepsilon_{\mathrm{read}}\le C[n^{-710}+\sqrt{B_{\mathrm{mean}}n^{-100}/\sigma}],
\qquad B_{\mathrm{mean}}\le Cn^2.
\end{equation}
All finite round, clip, invalid and overflow codes remain. The hard-step comparison is introduced after the smooth whole-clock comparison. It may lose pointer kinetic regularity and is subsequently used only under common free evolution and a suffix that never reads that old pointer. It supplies no actual energy graph.

This proves a full spin/App/$R$ comparison. The final bounded-Lipschitz macro marginal approaches the point density-flow prediction with its finite Gaussian readout. It is not a total-variation limit to a Dirac pointer state or a general microscopic classical model.

\setcounter{equation}{0}
\renewcommand{\theequation}{E\arabic{equation}}
\section*{Appendix E: Bell correlations on the same labels}
In the active phases $0,\pi$, a known local rotation maps the prepulse state to all-up. With the common phase retained,
\begin{equation}
|\psi(t)\rangle=e^{-i(J+2)t}\left[c|\mathrm{vac}\rangle+s\sum_{v=1}^4f_v(t)|v\rangle\right],
\quad f(t)=e^{2itL_{\mathrm{path}}}e_3,
\end{equation}
where $c=\cos(\theta/2)$, $s=\sin(\theta/2)$, and the path has weights $(1,J,1)$ in the order $(\tau_i,\sigma_i,\sigma_j,\tau_j)$. Define $P_A=|f_1|^2+|f_2|^2$, $P_B=|f_3|^2+|f_4|^2$. These are normalized one-flip masses, $P_A+P_B=1$, not masses already multiplied by $s^2$.

The local vacuum/one-flip amplitude matrix is
\begin{equation}
\begin{pmatrix}c&s\sqrt{P_B}\\s\sqrt{P_A}&0\end{pmatrix},
\end{equation}
so concurrence is $2s^2\sqrt{P_AP_B}$ and the fixed-$t_*$ Schmidt analyzers give
\begin{equation}
S_{\mathrm{active}}(t_*)=2\sqrt{1+4\sin^4(\theta/2)P_A(t_*)P_B(t_*)}=2+\eta_B>2.
\end{equation}
These extend to actual four-dimensional local instruments. The inactive $\pm Y$ phases produce the declared $+1$ defaults with score two and remain in the four-phase average.

Only after the coarse records are copied does one independent uniform global setting pair $x,y$ enter the two local banks. Past records and preparation flags may be shared; no bank receives the other fresh setting. Each returns $K$ bits. For
\begin{equation}
V=4(-1)^{xy}K^{-1}\sum_\ell a_\ell b_\ell,\qquad |V|\le4,
\end{equation}
the local bound is two even when either party processes all labels jointly. Every component has that bound, hence so does its mean. The reference score is $2+\eta_B/2$, while an actual bad branch has score at least $-4$. Thus
\begin{equation}
\mathbb EV\ge(2+\eta_B/2)p_s-4(1-p_s)-4\varepsilon_{\mathrm{joint}},\qquad
\liminf_n\mathbb EV\ge2+\eta_B/2.
\end{equation}
No phase, native outcome, field vacuum or apparatus success selects the sample. Finite positive gates implement the local analyzers with their prescribed calibration. The world remains on: its short pulse/readout windows cost $Cn^{-32}$ in the statistical local-channel comparison. No spacelike or device-independent implementation is claimed.

Figure~2 varies the reference analyzer calibration with its plotted time. The physical experiment uses its one fixed $t_*$ and analyzer set. Its collective and Bell records use the same material and memories. A separate fresh-preparation bath-return experiment is not part of this history.

\setcounter{equation}{0}
\renewcommand{\theequation}{F\arabic{equation}}
\section*{Appendix F: The complete error and resource conditions}
For the fixed library, the joint channel error has a conservative positive account
\begin{align}
\varepsilon_{\mathrm{joint}}\le C[&\varepsilon_{\mathrm{native}}+\eta_{\mathrm{in}}+n^{-4}
 +n^{-7}\tau+n^{-21/2}\tau^2+\varepsilon_{\mathrm{clk}}+\varepsilon_{\mathrm{on}}+\varepsilon_{\mathrm{cal}}\\
&+m_{\mathrm{obs}}/(\sigma\sqrt K)+C_md_g/\sigma
 +C_mn^{-710}+C_m\sqrt{B_{\mathrm{mean}}n^{-100}/\sigma}+\varepsilon_{\mathrm{time}}]\to0.
\end{align}
The native term includes the $2\varepsilon_{\mathrm{init}}$ glue, source, actual native/read, finite processor, dirty-helper and encoding fees. The all-trial point-record prediction separately pays $C(1-p_s)$. The label contribution $n^{-15/16}$ is one term, not a total error rate. Invalid quantum feedback is never assigned a hot classical law.

Whole-program free Gaussian clocks have their prescribed means, covariances, chirps and nonzero early/late tails. One choice uses $v_C=n^{3000}$, $m_C=n^{10000}$, width $n^{-20}$ and $T_{\mathrm{abs}}\le n^{18}$, yielding $\varepsilon_{\mathrm{clk}}\le Cn^{-990}$. The right clock factor is justified on the smooth program before introducing hard copies; actual recoil is retained on the left. Source-on and driver work are charged over the true on-time.

The stated positive constitutive, Gaussian, spin/App and finite Born primitives give
\begin{equation}
\mathrm{energy, work, clock, terms, precision}\le C[R_0(n)+n^{20000}].
\end{equation}
$R_0$ contains the actual source, native, bank, calibration and supplied-state certificates. Their costs are not inferred from trace error. Native pointer-momentum conservation, additive switching work and bounded interaction pinching avoid a coercivity factor multiplied once per gate. Common infinite free energies are factored exactly; unknown states and spectators are not called finite-cost supplies. Every repeat pays a complete fresh world, bath, source and apparatus preparation.

The bath length remains finite, with initial slip, free noise, source/purifier phases and nonzero image terms retained in the formation proof. On this clock and short body-observation window, the controlled endpoint comparison precedes the first reflected response. The later readout has its own scales and much larger trial/preparation account in the companion return manuscript. It grants no post-return matching or Bell experiment to this one.

\end{document}